\documentclass[a4paper,11pt]{article}
\usepackage{pos}
\usepackage[backend=biber,url=false,doi=false,hyperref,style=numeric-comp,sorting=none,bibencoding=utf8,maxnames=2,uniquename=init,giveninits=true]{biblatex}
\usepackage{braket}
\usepackage{amsmath}
\usepackage{physics}
\usepackage{booktabs}
\usepackage{soul}
\usepackage{multirow}
\usepackage{bbm} 
\usepackage{bm} 
\usepackage{tikz}
\usetikzlibrary{decorations.markings}
\usetikzlibrary{svg.path,calc}
\usetikzlibrary{patterns}
\usetikzlibrary{svg.path}
\usetikzlibrary{shapes,fit}
\usetikzlibrary{shapes.misc}

\AtEveryBibitem{
\clearfield{title}
}

\renewcommand{\eval}[1]{\langle#1\rangle}
\newcommand{\Ns}{N_{\mathrm{s}}}
\newcommand{\Nf}{N_{\mathrm{f}}}

\title{Variance reduction strategies for lattice QCD}

\author*{Tim Harris}

\affiliation{Institute for Theoretical Physics,\\
Department of Physics, ETH Z\"urich, Switzerland.}

\emailAdd{harrist@phys.ethz.ch}

\abstract{
A significant component of the cost of making predictions from lattice QCD
stems from the computation of correlation functions on a given ensemble of
gauge fields.
This cost depends on the observable of interest and the details of its
representation, including any approximation needed to estimate it.
Moreover, the variance of such estimators may depend strongly on physical and
kinematical parameters such as the lattice spacing, volume or separation, which
gives an important insight into the costs of reaching the relevant physical
limits.
In these proceedings, I review some observables involving quark propagators,
including both quark-line connected and disconnected Wick contractions, and
discuss variance-reduction schemes based on decompositions of the quark
propagators.
Such strategies have already proven useful for precision physics observables
and in future may help reduce the computational cost of reaching large
volumes.
}

\FullConference{The 42nd International Symposium on Lattice Field Theory (LATTICE2025)\\
2-8 November 2025\\
Tata Institute of Fundamental Research, Mumbai, India\\}

\begin{document}
\maketitle

\section{Introduction}
\label{sec:introduction}

Signal-to-noise problems are ubiquitous in Monte Carlo simulations of
lattice field theories.
In particular, deteriorating signal-to-noise ratios in the various numerical
limits required to make physics predictions may significantly hinder the
predictive power of lattice computations.
Let us immediately consider a concrete example, pure $\mathrm{SU}(3)$ gauge
theory regulated on a toroidal hypercubic lattice with lattice spacing $a$ and
side length $L$ defined by the Wilson action~\cite{Wilson:1974sk}
\begin{align}
    S_\mathrm{g} = \frac{2}{g_0^2}a^4\sum_{x}s(x),\qquad
    s(x) &= \frac{1}{a^4}\sum_{\mu<\nu} \Re\Tr{1 - P_{\mu\nu}(x)}
    \label{eq:wilson}
\end{align}
where $P_{\mu\nu}(x) =
U_\mu(x)U_\nu(x+a\hat\mu)U^\dagger_\mu(x+a\hat\nu)U^\dagger_\nu(x)$ is the
plaquette based at site $x$ oriented in the $\mu\nu$ plane.
After the subtraction of its vacuum expectation value, the product of the
density $s(x)$ and the (lattice) beta function $\mathrm dg_0^{-2}/\mathrm d\ln
a$ is a renormalized field~\cite{Boyd:1996bx} which we denote $\phi(x)$, and
therefore its connected Euclidean time-correlation function
\begin{align}
    C(x_0-y_0) &= a^3\sum_{\vec x} \Big[\eval{\phi(x)\phi(y)} -
    \eval{\phi(x)}\eval{\phi(y)}\Big]
    \label{eq:correlator}
\end{align}
has a finite value for non-zero separations, $x_0-y_0>0$, in the continuum and
infinite volume limit, and moreover is exponentially suppressed at large
separations due to the mass gap $m$ of the theory.

A standard Monte Carlo simulation then yields an estimate for each of the path
integrals in eq.~\eqref{eq:correlator} via the sample mean, for example
\begin{align}
    \eval{\phi(x)} \approx \frac{1}{N}\sum_{i=1}^N \phi^i(x),
    \label{eq:mean}
\end{align}
thanks to the law of large numbers, where the superscript indicates the
{density} is evaluated on the $i$th field sample.
Assuming, then, that the samples are independent and identically distributed,
the variance of the plug-in estimator is found to be
\begin{align}
    \frac{\sigma^2}{N} &= \frac{1}{N}\Big[a^6\sum_{\vec x,x'}\delta_{x_0,x_0'}
    \eval{\phi(x)\phi(x')}_\mathrm{c}\eval{\phi(y)\phi(y)}_\mathrm{c} + 
    \mathrm O(\mathrm e^{-m(x_0-y_0)})\Big] + \ldots,
    \label{eq:lln}
\end{align}
where the ellipses denote terms higher order in $1/N$.
The utility of this formula is that the statistical uncertainty is completely
determined by the physics of the underlying system and therefore much can be
learned about the signal-to-noise ratio in the interesting limits.

One immediate consequence of the preceding formula is that, although the
correlation function is well-behaved in the limits $a\rightarrow0$,
$L\rightarrow\infty$ and $x_0-y_0\rightarrow\infty$, the same is not true for
the variance of its Monte Carlo estimator.
Indeed, one may expect the following asymptotic behaviour for
the signal-to-noise ratio in those same limits
\begin{align}
    \frac{C(x_0-y_0)}{\sqrt{\sigma^2/N}} \sim  a^5
    \sqrt\frac{a^3}{L^3}
        \mathrm e^{-m (x_0-y_0)}\sqrt{N},
    \label{eq:stn}
\end{align}
by considering the dominant contributions coming from the points where the
fields coincide and these products mix with the vacuum, see
Ref.~\cite{Altenkort:2021jbk} for another worked example.
The severity of such problems has long been
appreciated~\cite{Parisi:1983ae,Lepage:1989hd}, and can be (partially) dealt
with.
In fact the simplest improvement is to average over the volume, and we content
ourselves with this modest goal in the rest of these proceedings.

\subsection{Translation averaging}
\label{sub:translation_averaging}
As is well-known (e.g.~Ref.~\cite{Luscher:2010ae} for an exposition), assuming
local translation invariance, an improved estimator for the correlation
function is obtained by averaging the product of fields over a region of the
lattice, for example on a time-slice
\begin{align}
    C_\mathrm{vol}(x_0-y_0) &= \frac{a^3}{L^3}\sum_{\vec y}a^3\sum_{\vec x} 
    \Big[\eval{\phi(x)\phi(y)} - \eval{\phi(x)}\eval{\phi(y)}\Big]
    \label{eq:vol}
\end{align}
which has the beneficial effect of suppressing the variance of the
corresponding estimator by a factor $(a/L)^3$ and therefore enhancing the
signal-to-noise ratio
\begin{align}
    \frac{C(x_0-y_0)}{\sqrt{\sigma^2_\mathrm{vol}/{N}}} 
    \sim a^5 \mathrm e^{-m (x_0-y_0)}\sqrt{N} 
    \label{eq:volvar}
\end{align}
Even better, is to avoid integrating the fields over large transverse
separations in the first place (see
Refs.~\cite{Bali:2009dz,Blum:2015gfa,Liu:2017man,Meyer:2017hjv,Luscher:2017cjh,Giusti:2018cmp,Bruno:2023vhs}
for some examples), and say, cutting off the contribution from the tail after
some distance $R$,
\begin{align}
    C_{\mathrm{vol},R}(x_0-y_0) &= \frac{a^3}{L^3}\sum_{\vec y}
    a^3\sum_{\vec x} \theta(R^2-|\vec x-\vec y|^2)
    \Big[\eval{\phi(x)\phi(y)} - \eval{\phi(x)}\eval{\phi(y)}\Big]
    \label{eq:cut}
\end{align}
giving rise to a signal-to-noise ratio which even increases with the lattice size
\begin{align}
    \frac{C(x_0-y_0)}{\sqrt{{\sigma^2_{\mathrm{vol},R}}/N}} \sim 
    a^5\sqrt{\frac{L^3}{R^3}}\mathrm e^{-m (x_0-y_0)}
    \sqrt{N}.
    \label{eq:cutvar}
\end{align}
That estimator has the attractive feature of improving the statistical and
systematic error associated to finite-volume effects in one go, if the regime
can be reached for which $L/R$ is sufficiently large that the tail
contribution is under control.
Of course, the formulae presented so far are applicable only to the asymptotic
regions of parameter space: the precise scaling in intermediate regimes may be
different and motivate the generalization of the averages to other domains
which may or may not be contiguous, as well as the obvious generalization to
averaging over the time extent.

Other $n$-point functions can be analysed in much the same spirit as, almost
universally, the problematic features arise from the fact that the variance is
not a fully-connected correlation function and disconnected components are
practically impossible to avoid.
An important exception to this rule will be discussed in the next section.
Certain advanced strategies may mitigate the poor scaling in the other
variables, by using for example multi-level integration discussed in
Sec.~\ref{sec:multi-level}.
However, even implementing the most simple strategy, namely translation
averaging, is not always easy.

\subsection{Translation averaging in QCD}
\label{sub:cost}
So far, we have neglected to discuss the cost of the three different estimators
presented.
Happily, due to the fact that the products of local fields discussed so far
factorize, the translation average of the previous section can be evaluated
with the help of the convolution theorem and so its cost grows only like
$V\ln V$, where $V=(L/a)^3$ in the case above, i.e. not too much more than the
original estimator.
In general, however, the cost to achieve a reduced variance may not be so
favourable and one should keep in mind that the relevant metric to be optimized
is, for example the computational cost for a fixed precision,
$\varepsilon=\sqrt{\sigma^2/N}$, which is
\begin{align}
    \textrm{total cost} &= N\times\textrm{cost per field}
    = \varepsilon^{-2}[\sigma^2\times\textrm{cost per field}].
    \label{total_cost}
\end{align}
A good variance reduction strategy minimizes the product in the brackets,
i.e.~the effective cost.

\begin{figure}[t]
    \centering
    \begin{tikzpicture}[scale=0.5,remember picture,every text node part/.style={align=left}]
    \begin{scope}[thick, decoration = {
            markings,
            mark=at position 0.5 with {\arrow{>}}}]
    \node at (0,0) {};
    \begin{scope}[yshift=-1.5cm]
        \draw[line width=0.2em] (2.4,0) -- (2.4,8.4) node[above] {$x_0$};
        \draw[line width=0.2em] (6.3,0) -- (6.3,8.4) node[above] {$y_0$};
        \draw[thick,xstep=0.3,ystep=0.3] (0,0)  grid (8.4,8.4);
        \draw[very thick,->] (-0.5,-0.5) -- (-0.5,2) node [above] {$\hat i$};
        \draw[very thick,->] (-0.5,-0.5) -- (2,-0.5) node [right] {$\hat 0$};
        \node[fill=white,draw] at (1.1,7.5) {{\color{red}$\sigma^2_{\mathrm{conn}}$}};
        \shade[inner color=red,  opacity = 1.00, draw, thick] (2.4,2.1) node (p1) {} circle (0.25);
        \shade[inner color=red,  opacity = 1.00, draw, thick] (2.4,3.1) node (p3) {} circle (0.25);
        \shade[inner color=red,  opacity = 1.00, draw, thick] (6.3,6.1) node (p2) {} circle (0.25);
        \shade[inner color=red,  opacity = 1.00, draw, thick] (6.3,7.1) node (p4) {} circle (0.25);
        \node[ellipse,fit={(p1)(p2)},inner color=red, minimum height=1cm,opacity=0.25,rounded corners] {};
        \node[ellipse,fit={(p3)(p4)},inner color=red, minimum height=1cm,opacity=0.25,rounded corners] {};
        \draw[very thick,postaction = {decorate}] (p1) .. controls ++(30:2) and ++(60:-2) .. (p2);
        \draw[very thick,postaction = {decorate}] (p2) -- (p1);
        \draw[very thick,postaction = {decorate}] (p3) -- (p4);
        \draw[very thick,postaction = {decorate}] (p4) .. controls ++(30:-2) and ++(60:2) .. (p3);
    \end{scope}
    \begin{scope}[yshift=-1.5cm,xshift=12.5cm]
        \draw[line width=0.2em] (2.4,0) -- (2.4,8.4) node[above] {$x_0$};
        \draw[line width=0.2em] (6.3,0) -- (6.3,8.4) node[above] {$y_0$};
        \draw[thick,xstep=0.3,ystep=0.3] (0,0)  grid (8.4,8.4);
        \shade[inner color=blue,  opacity = 1.00, draw, thick] (2.4,2.1) node (p1) {} circle (0.25);
        \shade[inner color=blue,  opacity = 1.00, draw, thick] (2.4,3.1) node (p3) {} circle (0.25);
        \shade[inner color=blue,  opacity = 1.00, draw, thick] (6.3,6.1) node (p2) {} circle (0.25);
        \shade[inner color=blue,  opacity = 1.00, draw, thick] (6.3,7.1) node (p4) {} circle (0.25);
        \node[rectangle,fit={(p2)(p4)},inner color=blue, minimum height=1cm,opacity=0.5,rounded corners] {};
        \node[rectangle,fit={(p1)(p3)},inner color=blue, minimum height=1cm,opacity=0.5,rounded corners] {};
        \draw[very thick,->] (-0.5,-0.5) -- (-0.5,2) node [above] {{$\hat i$}};
        \draw[very thick,->] (-0.5,-0.5) -- (2,-0.5) node [right] { $\hat 0$};
        \draw[very thick,postaction = {decorate}] (p1) .. controls (3.7,3.1) and (3.7,1.1) .. (p1);
        \draw[very thick,postaction = {decorate}] (p2) .. controls (5.2,7.1) and (5.2,5.1) .. (p2);
        \draw[very thick,postaction = {decorate}] (p3) .. controls (3.7,4.1) and (3.7,2.1) .. (p3);
        \draw[very thick,postaction = {decorate}] (p4) .. controls (5.2,8.1) and (5.2,6.1) .. (p4);
        \node[fill=white,draw] at (1.1,7.5) {{\color{blue}$\sigma^2_{\mathrm{disc}}$}};
    \end{scope}
    \end{scope}
\end{tikzpicture}
    \caption{Illustration of the Wick contractions which appear in the
    variances for a quark-line connected (left) and disconnected (right)
    primary observable.}
    \label{fig:variances}
\end{figure}
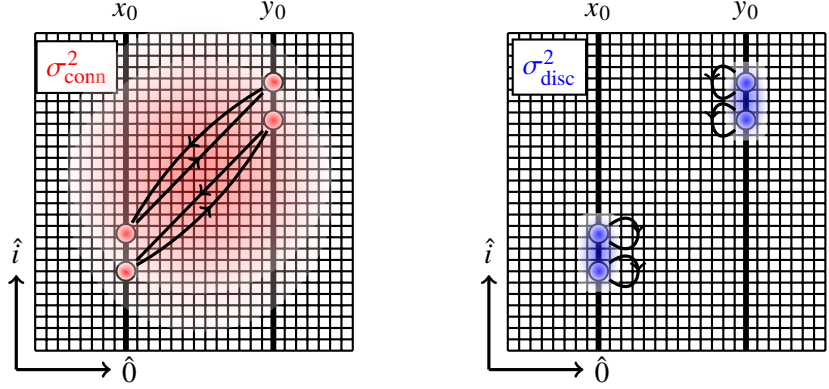

In QCD, in the typical representation of the path integral, the local quark
fields are integrated out by hand, leaving observables written in terms of
products of quark propagators of flavour $f$,
$S_f(x,y)=\eval{\psi_f(x)\bar\psi_f(y)}_\mathrm{F}$, the Greens function for
the lattice Dirac operator $D_f$, and consequently no longer necessarily
factorize.
To implement translation averaging exactly, the quark propagator should be
computed for all $x$ and $y$ which is a problem with $\mathrm O(V^2)$
complexity and thus the effective cost grows with the volume.
In the sense of the above, then, we seek estimators which minimize the
effective cost by approximating the translation average, and ideally one for
which it scales favourably with the physical volume.

As a prototypical set of observables, we consider the Euclidean-time
correlation function of bilinears of fermion fields, and in particular the
electromagnetic current, $j_\mu=\sum_f Q_f\bar\psi_f\gamma_\mu\psi_f$, where
$Q_f$ is the electric charge of the quark flavour $f$, which is an especially
important operator as its hadronic matrix elements parameterize photon-hadron
interactions in the Standard Model~\cite{Bernecker_2011}.

The bare spatial current correlator
\begin{align}
    G(x_0,y_0) &= \frac{a^3}{L^3}\sum_{\vec y} a^3\sum_{\vec x}\sum_{f,g}
    Q_fQ_g \eval{T^{(1)}_{k,f}(x)T^{(1)}_{k,g}(y)-T^{(2)}_{k,fg}(x,y)\delta_{fg} }
    \label{eq:curr_correlator}
\end{align}
where the brackets denote the expectation with respect to the effective
distribution $\det{D_f}^{\Nf}\mathrm e^{-S_\mathrm{g}}$, can be written in terms
of two traces, the single-propagator trace
\begin{align}
    T_{\mu,f}^{(1)}(x) = \tr{\gamma_\mu S_f(x,x)},
    \label{eq:trace1}
\end{align}
which is one factor of the quark-line disconnected graph,
and the trace of two quark propagators
\begin{align}
    T_{\mu,fg}^{(2)}(x,y) = 
    \tr{\gamma_\mu S_f(x,y)\gamma_\mu S_g(y,x)},
    \label{eq:trace2}
\end{align}
which we refer to as the quark-line connected Wick contraction.

The variances are now defined in terms of the simulated system, and typically
cannot be represented in terms of correlation functions of the original
fields, so the analysis of the pure gauge theory must be extended to a
partially-quenched set-up (e.g.~Refs.~\cite{DellaMorte:2005nwx,DellaMorte:2010aq}).
Interestingly, the consequences of operating with the non-local Wick
contractions as observables is not totally detrimental: the explicit factors
of the quark propagator, which is naturally small $\mathrm O(\mathrm
e^{-m_\pi\abs{x-y}/2})$ independently of the field (its variance is the pion
propagator), suppresses the variance of
the quark-line connected contractions with physical separations, and even
forbids mixing of the effective vertex with the vacuum state, see
Fig.~\ref{fig:variances} for a sketch.
That is, the variance, too, is a fully-connected graph and suffers no
power-law divergences in the continuum limit while the contributions with
large transverse separations are naturally suppressed.
Quark-line disconnected diagrams, which arise whenever singlet operators are
considered, on the other hand, are presumably afflicted by the issues
described in previous section~\cite{Giusti:2019kff,Harris:2024dky}, although
empircal evidence may suggest the asymptotic regime is not
reached~\cite{Altherr:2025juv}.
This has important consequences for the method of
insertions~\cite{RC:2025zoa}, for example the computation of isospin-breaking
effects where disconnected diagrams necessarily arise in the Rome-123
approach~\cite{deDivitiis:2013xla}.
Therefore, it is useful to distinguish the two cases, and in practice it
points to the fact that schemes with a good representation of the quark
propagator at small distances for the disconnected graphs and at large
distances for the connected graphs are essential.

\subsection{Stochastic estimates for translation averages}
\label{sec:stochastic_estimates}
As outlined in the previous section, constructing the translation average
explicitly for observables built from quark propagators is prohibitively
expensive and even and has an effective cost which increases with the
physical volume.
That is, it would be more favourable to increase the number of configurations
than to increase the volume from the point of view of the statistical
precision.
Instead, we can opt to estimate the sums over the coordinates
stochastically, and in this work we opt to make use of generalizations of the
simple Hutchinson estimator for the trace of a
matrix~\cite{Bitar:1988bb,Hutch1990,Dong:1993pk,deDivitiis:1996qx,Michael:1998sg},
defined by computing quadratures with a set of dimensionless random fields
$\eta_i(x)$ whose
independent components have zero mean and unit variance\footnote{%
One may wonder whether the choice of the white noise distribution has an
effect on the variance~\cite{BERNARDSON1994256}, but in fact, it appears to
play a minor role in practice, and choosing Gaussian fields simplifies
the analysis of the variance.}
and support on a
time-slice $x_0$
\begin{align}
    \label{eq:hutch1}
    {a^3} \sum_{\vec x} T_{\mu,f}^{(1)}(x) &\approx \frac{1}{\Ns}
        \sum_{i=1}^{\Ns} a^3\sum_{\vec x}\eta^\dagger_i(x)\gamma_\mu(S_f\eta_i)(x), \\
    {a^6} \sum_{\vec x,\vec y} T_{\mu,fg}^{(2)}(x,y) &\approx \frac{1}{\Ns}
        \sum_{i=1}^{\Ns} a^3\sum_{\vec y}
        (\eta^\dagger_i\gamma_\mu S_f)(y)\gamma_\mu(S_g\eta_i)(y).
    \label{eq:hutch2}
\end{align}
Indeed, with that choice, the variance of the single-propagator trace can be
worked out as~\cite{Giusti:2019kff}
\begin{align}
    \nonumber
    \sigma^2 &= a^6\sum_{\vec x,x'}\delta_{x_0,x_0'}\Big[
        \eval{T_{\mu,f}^{(1)}(x)T_{\mu,f}^{(1)*}(x')}
        - \eval{T_{\mu,f}^{(1)}(x)}\eval{T_{\mu,f}^{(1)*}(x')}\\
        & \qquad\qquad\qquad + \frac{1}{\Ns}\eval{T^{(2)}_{5,ff}(x,x')}\Big].
    \label{eq:variance}
\end{align}
The first line corresponds to the variance associated with the gauge fields
(often referred to as the gauge variance $\sigma^2_\mathrm{g}$) while the term
on the second line is due to the auxiliary random fields and, as expected
vanishes in the limit of a large number of fields, $\Ns\rightarrow\infty$.
This illustrates that, although the estimator provides an unbiased estimate of
the translation average, and its cost depends on the number of auxiliary
random (source) fields and not explicitly on the volume, the variance receives
an extra contribution, which in the case of the vector current is large.

\begin{figure}[t]
    \centering
    \includegraphics{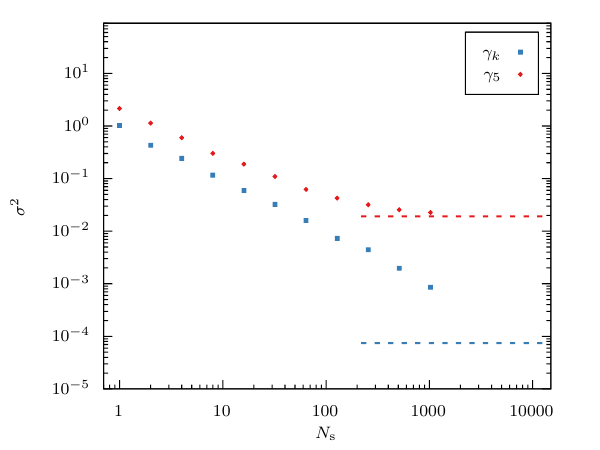}
    \caption{The variance of the Hutchinson estimator for the
    single-propagator trace $T^{(1)}_{\mu,f}$ for $\mu=k$ (blue squares) and
    $\mu=5$ (red circles).
    The dashed lines indicate the gauge variance which differs by many orders
    of magnitude for the two cases unlike the stochastic variance which is
    identical if the reality of observable is ignored.}
    \label{fig:fig2-vector-ps}
\end{figure}

This is visible from simulations performed with $\mathrm O(a)$-improved Wilson
fermions and a pion mass of $m_\pi\approx270\,\mathrm{MeV}$ and $m_\pi
L\approx 4$ (described in detail in Sec.~\ref{sec:freq-split}) depicted in
Fig.~\ref{fig:fig2-vector-ps} which shows that the additional contribution is
enhanced by several orders of magnitude compared to the gauge variance, indicated
with the horizontal line. 
Some indications of this problematic behaviour from the formula of the
variance come from the fact that the additional term corresponds to the
integrated pion propagator independently of the gamma-matrix in the original
estimator, which also has a non-zero value at leading order in perturbation
theory, unlike the gauge variance.

A similar exercise reveals an analogous situation for the trace of two
propagators for the vector current~\cite{Gruber:2024czx,Luscher:2010ae}.
That is, in practice such simple stochastic estimators introduce large
additional fluctuations and which greatly increase the effective cost even if
they solve the volume-scaling problem in principle.
The goal then is to find improved estimators based on decompositions of the
quark propagators which allow us to efficiently reduce the effective cost by
coming up with good and inexpensive approximations to the propagator either at
short- or long-distances.
If the approximation is cheap, then it can be estimated precisely by sampling
frequently, and if the approximation is good then the correction will be
suppressed and consequently also have a small error.


\newcommand{\Nv}{N_\mathrm{v}}

\section{Multigrid low mode averaging}
\label{sec:mg-lma}

In order to suppress the variance for the stochastic estimator for the
quark-line connected diagram corresponding to the trace of two quark
propagators of eq.~\eqref{eq:trace2}, we seek a good approximate
representation of the quark propagator at long distances.
To this end, we utilize the concept of deflation of the lattice Dirac operator
$D$, by introducing a subspace spanned by $\Nv$ global quark fields
$\phi_1(x),\ldots,\phi_{\Nv}(x)$.
In this section, we drop the flavour index $f$ for brevity.
Associated with this deflation subspace, one can define a projector $R$
defined through its action on a quark field $\psi$ by
$(R\psi)_i=(\phi_i,\psi)$, and an associated little Dirac operator $\hat
Q=RQR^\dagger$, which acts in the subspace, sometimes known as the Galerkin
coarsening, of the Hermitian Dirac operator $Q=\gamma_5 D$.

If the observable is dominated by propagation in this subspace, then one may
expect the operator $R^\dagger\hat Q^{-1}R$ to furnish us with a good
approximation to it.
We can form a decomposition by addition and subtraction
\begin{align}
    S(x,y) &= \underbrace{\{S(x,y) -  (R^\dagger\hat
    Q^{-1}R\gamma_5)(x,y)\}}_{\mathcal S_0(x,y)}
    + \underbrace{(R^\dagger\hat Q^{-1} R\gamma_5)(x,y)}_{\mathcal S_1(x,y)}
    \label{eq:lma}
\end{align}
where the term in braces corresponds to the Green's function for the deflated
system, and is related to the coarse-grid correction operator as it serves to
correct the approximant given by the second term.
If the approximation is a good one, one may expect that the first term along
with its variance will be highly suppressed.
Inserting the decomposition of the quark propagator in the quark-line
connected diagram, we arrive at the decomposition
\begin{align}
    \nonumber
    T^{(2)}_\mu(x,y) &=   \tr{\gamma_\mu \mathcal S_0(x,y)\gamma_\mu \mathcal S_0(y,x)} 
                        + \tr{\gamma_\mu \mathcal S_0(x,y)\gamma_\mu \mathcal S_1(y,x)} 
                        + \tr{\gamma_\mu \mathcal S_1(x,y)\gamma_\mu \mathcal S_0(y,x)} \\
                        &\qquad\qquad+ \tr{\gamma_\mu \mathcal
                        S_1(x,y)\gamma_\mu \mathcal S_1(y,x)} 
    \label{eq:trace2-lma}
\end{align}
where each term in the first line requires the expensive deflated operator to
be computed, while the term on the second line requires only the inversion of
the little operator.

If $\Nv$ is sufficiently small, then the little operator is clearly cheap
to apply, while if it is furthermore well-conditioned, we may expect that it
is also inverted without too much effort.
In fact, if one takes the deflation subspace to be spanned by a few exact
eigenmodes of $Q$ with small in magnitude eigenvalues, which is called
low-mode averaging~\cite{Giusti_2004,DeGrand_2004,Foley:2005ac}, then the
inversion of the little operator is indeed trivial.
In order to suppress the variance in the costly remainder term, then, a small
number of low modes needs to provide a good approximation to the quark
propagator at long distances.
Unfortunately, the required number of low-modes for a constant suppression of
the variance grows with the physical volume, which is essentially a
consequence of the increasing density described by the Banks-Casher
relation~\cite{banks1980}.
In the following, we circumvent the issue by exploiting the property of the
local coherence of the low modes, as suggested already in
Ref.~\cite{Luescher2007}.

\newcommand{\Nc}{N_\mathrm{c}}
\newcommand{\Nb}{N_\mathrm{b}}

\subsection{Local coherence and deflation}
\label{sec:local-coherence}
The low quark modes exhibit a property known as local coherence (or
weak-approximation) which is that the support of such fields on a small
domain, of linear size $b=0.25-0.5\,\mathrm{fm}$, span a small
subspace compared with their total number.
This suggests using as a basis a few low modes projected to small
domains to create a subspace whose size grows with the volume without any
extra cost.
The use of such blocked low modes has been instrumental in the definition of
efficient deflated or multigrid preconditioned
solvers~\cite{Luescher2007,babich2010,Frommer:2013}, and is crucial aspect of
forming a local averaging (or aggregation) in gauge theories, and has also been
applied to compression of fields~\cite{Clark_2018}.
Once the lowest level averaging is defined, the extension to a hierarchical
scheme is fairly trivial by simply combining the blocks, which gives a flexible
approach to tuning the effective cost~\cite{Gruber:2024cos}.
A similar scheme to what follows has been proposed in
Ref.~\cite{Frommer:2025wew}.

\begin{figure}[t]
    \centering
    \includegraphics[scale=0.5]{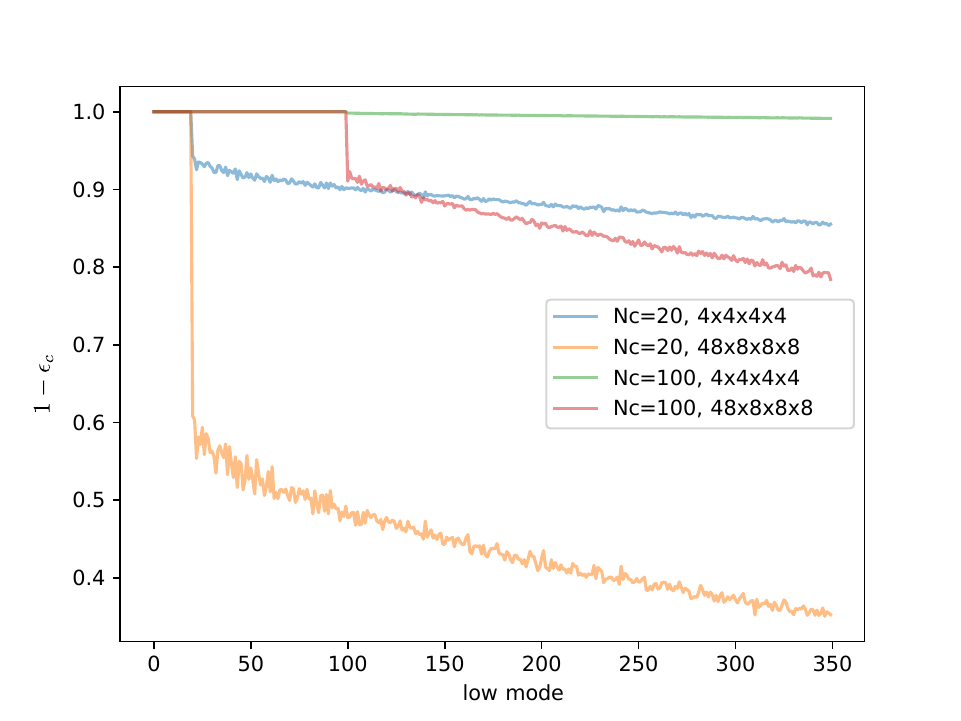}
    \caption{One minus the deficit of the lowest modes in subspaces built from
        just a few $\Nc=20$ (blue or yellow) or 100 (green or red) low modes.
        Decreasing the block sizes (indicated in lattice units) increases the
        subspace size and reduces the deficits further.}
    \label{fig:deficit}
\end{figure}

Following Ref.~\cite{Gruber:2024cos}, we imagine the $\Nv=2\Nc\Nb$ block low
modes are constructed from $\Nc$ exact low modes of the Dirac operator,
$\phi^{B,\pm}_{c}(x)=\theta_B(x)(1\pm\gamma_5)\phi_c(x)$, where $\theta_B(x)$
is the characteristic function of the block labelled by $B=1\ldots\Nb$.
After orthonormalizing, the resulting restricted fields, now called coarse
fields because they carry geometric information, have components
\begin{align}
    (R\psi)^\pm_c(x_B) =  (\phi^{B,\pm}_c,\psi)
    \label{eq:coarse_fields}
\end{align}
where $x_B$ are the coordinates of the block $B$, and have a structure
reminiscent of the original quark field.
The alignment of a field $\psi$ with the subspace is large if its deficit 
\begin{align}
    \epsilon_\psi = \norm{\psi - R^\dagger R\psi}/\norm{\psi}
    \label{eq:deficit}
\end{align}
is small.
Fig.~\ref{fig:deficit} shows that the deficits of first few hundred exact low
modes are small for subspaces built with just a small number $\Nc=20,100$ on
an ensemble with $m_\pi L\approx 4$ as described in the next subsection.
This indicates that expanding the dimension of the subspace by the blocking
procedure should improve the representation of the quark propagator without
needing $\mathrm{O}(V)$ low modes to begin with.
In the following, we investigate empircally both conditions for an efficient
deflation variance reduction scheme, namely that the inversion of the little
operator is cheap and that the variance on the expensive correction term is
suppressed.

\subsection{Numerical investigations}
\label{sub:mg-lma-numerical}

\begin{table}[t]
    \centering
    \begin{tabular}{ccccccc}
        \toprule
        \multirow{2}{*}{{$m_\pi L$}} &
        \multirow{2}{*}{{estimator}} &
        \multicolumn{4}{c}{$\Ns$} &
    \multirow{2}{*}{$\frac{\textrm{measured cost}}{\textrm{Hutchinson $\Ns=1$}}$}
        \\
        \cmidrule(lr){3-6}
                           & & L0   & L1         & L2      & L3              &                        \\
        \midrule
        \multirow{2}{*}{2.9} & Hutchinson & 1024 & -     & -     & - & {$1024$} \\
                             & LMA        & 16   & exact & -     & - & {$16$}   \\
                             & MG LMA     & 1    & 16    & exact & - & {$1.4$}  \\
        \midrule
        \multirow{2}{*}{4.3} & Hutchinson & 2048 & -    & -     & -     & {$2048$} \\
                             &    LMA     & 1024 & 1024 & exact & -     & {$1024$} \\
                             & MG LMA     & 16   & 1024 & exact & -     & {$65.8$} \\
        \midrule
        \multirow{2}{*}{5.8} & Hutchinson & 4096 & -    & -     & -     & {$4096$} \\
                             &    LMA     & 2048 & exact& -     & -     & {$2048$}  \\
                             & MG LMA     & 1    & 16   & 1024  & exact & {$117$}  \\
        \bottomrule
    \end{tabular}%
    \caption{Indicative costs to achieve a fixed variance for three different
        lattice volumes for the undeflated estimator (Hutchinson), low-mode
        averaging (LMA) and multigrid low-mode averaging (MG LMA).
        All deflated schemes use $\Nc=50$
    exact low modes.}%
    \label{tab:conn}
\end{table}%
In this section, we report the variance of the stochastic estimator for the
deflated terms (the first line of eq.~\eqref{eq:trace2-lma}) and the
approximant (the second line) using $\Nf=2$ $\mathrm O(a)$-improved Wilson
fermions~\cite{Jansen:1998mx} for various lattice sizes $m_\pi L=2.9,4.3,5.8$ with
$m_\pi=270\,\mathrm{MeV}$ and $a=0.066\,\mathrm{fm}$.
The intermediate volume corresponds to the F7 ensemble generated by the CLS
consortium~\cite{Fritzsch:2012wq} while the others were generated with the
openQCD code~\cite{openqcd}.
The eigensolver from the PRIMME library~\cite{PRIMME} was used to determine
the low modes.
This allows us to investigate the hypothesis that a small number of low modes
$\Nc$ fixed for all lattice sizes is sufficient to suppress the variance as
long as the blocked low modes are used in the definition of the deflation
subspaces.

For the hierarchichal scheme, the propagator is decomposed telescopically into
$N_\ell$ levels where the propagators on levels $0$ and $N_\ell-1$ have a
similar form as the two-level case and on the intermediate levels
$0<l<N_\ell-1$ are given by
\begin{align}
    \mathcal S_l = \mathcal R^\dagger_l Q_l^{-1}
    \mathcal R_l\gamma_5 -
    \mathcal R^\dagger_{l+1} Q_{l+1}^{-1} \mathcal R_{l+1}\gamma_5.
    \label{eq:level}
\end{align}
where the little operator on level $l$ is defined by as $Q_l=\mathcal R_l
Q\mathcal R_l^\dagger$.
The restrictors $\mathcal R_l$ on coarser levels are just defined by
decompositions into larger blocks, which, if they nest, even allow a truly
recursive procedure to be constructed.

For the connected correlator, the $N_\ell^2$ terms are then estimated in
$N_\ell$ levels
\begin{align}
    G(x_0,y_0) &\approx \sum_{l=0}^{N_\ell-1}G_{\mathrm Ll}(x_0,y_0)
\end{align}
where we define each level as the sum of terms
\begin{align}
    G_{\mathrm Ll}(x_0,y_0) =
    \frac{1}{\Ns}\sum_{i=1}^{\Ns}
    a^3\sum_{\vec y}\sum_{
    \substack{m,n=0\\\min(m,n)=l}}^{N_\ell-1}
    (\eta_i^\dagger\gamma_k\mathcal S_m)(y)
    \gamma_k (\mathcal S_n\eta_i)(y)
    \label{eq:trace2-levels}
\end{align}
which collects all the terms which involve at least one inversion of the Dirac
operator on level $l$ and coarser levels.
Each level requires an inversion of the corresponding coarse or coarser
operators only.
Such a decomposition, assuming different sources on each correlator
level, has a variance
\begin{align}
    \sigma^2 = \sigma^2_\mathrm{g} +
    \frac{1}{N_\mathrm{L0}}\sigma^2_{\mathrm L0} + 
    \frac{1}{N_\mathrm{L1}}\sigma^2_{\mathrm L1} + \ldots,
    \label{eq:variance_levels}
\end{align}
where the variances due to the auxiliary fields can be written in terms of the
deflated propagators on that level.
All the parameters which define the estimators which include
$N_{\mathrm{L}l}$, the block sizes and the number of exact low modes $\Nc$
can be adjusted to optimize the effective cost.

In Tab.~\ref{tab:conn}, I report the cost of the Hutchinson estimator and
deflated schemes without blocking (LMA) and with blocking (MG LMA), either
three-levels with block sizes $8^4$ (in lattice units) and a
four-level scheme with an additional block size of $4^4$.
All of the deflated schemes use $\Nc=50$ exact low modes.
The cost to reach a fixed precision is in units of a cycle of the plain
Hutchinson estimator, which was implemented with spin-diagonal
sources~\cite{ETM:2008zte}.
The coarsest level has no blocking and thus can be computed exactly.
In all cases we observe that the little operators are at least as well
conditioned as the Dirac operator, and could be inverted using a standard GCR
solver~\cite{Gruber:2025eey}.
The results show that just a few sources are required for the expensive
deflated operator which translates into an large reduction in the effective
cost, remaining a factor 30 cheaper on the largest volume than without
deflation, without requiring a large number of low modes.
As expected, the advantage of LMA over the Hutchinson estimator quickly
evaporates as the physical volume is increased with a fixed number of low
modes, in constrast to MG LMA.

\begin{figure}[t]
    \centering
    \includegraphics[scale=0.65]{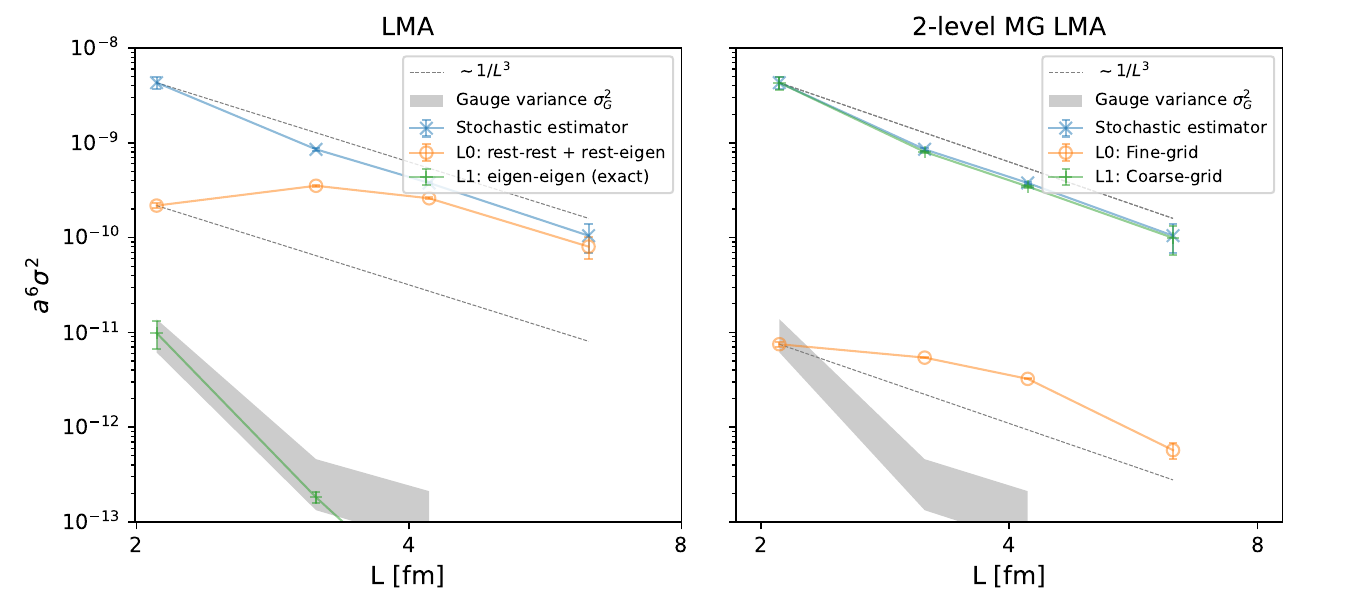}
    \caption{Volume dependence of the variance of the levels of the improved
        estimators without using block-projection of the low modes (LMA, left) and with
        block projection (MG LMA, right) for a separation $x_0-y_0\approx1.3\,\mathrm{fm}$.
    The variance of the undeflated estimator is also shown with blue points,
    while the grey band indicates an estimate of the gauge variance. All variances
    for the stochastic estimators are computed with $\Ns=1$ auxiliary field.}
    \label{fig:voldep}
\end{figure}

In Fig.~\ref{fig:voldep}, the volume dependence of the variance of each level
is shown for $\Ns=1$, using again $\Nc=50$ fixed for the deflated estimators,
for LMA (left) and MG LMA (right).
The grey band indicates the gauge variance.
The variance of the single (undeflated) Hutchinson estimator is shown in both
cases with the blue points and decreases with $L$, as expected, roughly like
$1/L^3$ (dashed lines).
While the variance of the deflated term (orange circles) is
suppressed on the smallest volume in LMA, it quickly saturates the variance
of the undeflated estimator, illustrating that the deflation loses efficiency
as the volume is increased but the number of modes is kept constant.
For MG LMA, when the low modes are blocked, the variance of the deflated term
does not saturate the variance of the Hutchinson estimator and decreases with
the volume.
This illustrates that the increase of the subspace size with the volume using
a fixed physical block size keeps this contribution small.

Evidently, such a scheme could be optimized further.
The implementation which was used to produce the cost estimates in
Tab.~\ref{tab:conn} relied on a fairly primitive iterative solver.
Using a true multi-grid preconditioned solver could improve the coarse-grid
solves and accelerate further the computation of the coarse-grid propagators.
The advantage of MG LMA is that just a few low modes are sufficient to deflate
the variance if they are subsequently blocked.
Whether using approximate low modes like those fields used in the set-up stage
of the deflated or multigrid solver would provide as good a basis remains to
be seen.
One interesting observation of this study is that, in order to fulfil the
criteria of a good approximation and a well-conditioned little system,
both chiralities must be kept in the subspace~\cite{Gruber:2025eey}.
Finally, the application to other quark-line connected diagrams with varying
number of light-quark propagators, such as static-light meson propagators with
one propagator, or baryon propagators with three, would be very interesting,
as would the application to fields which have been smoothened with a procedure
like distillation~\cite{HadronSpectrum:2009krc}.


\section{Frequency-splitting}
\label{sec:freq-split}

As motivated earlier, implementing translation averaging for quark-line
disconnected diagrams requires a good and cheap representation of the
propagator at short distances.
This can be achieved by noting that the quark propagator at short distances is
quite independent of the quark mass~\cite{Giusti:2019kff}.
Therefore, we can add and subtract the quark propagator with a larger mass
\begin{align}
    S_{f}(x,x) &= \underbrace{\{S_{f}(x,x) - S_{g}(x,x)\}}_{\mathcal S_0(x,x)} +
    \underbrace{S_{g}(x,x)}_{\mathcal S_1(x,x)},\qquad m_{f}< m_{g},
    \label{eq:mass-shift}
\end{align}
where we restrict immediately to the case where $x=y$.
When inserted into the single-propagator trace, this leads to a decomposition
of the trace into two levels, each of which can be evaluated independently.
Exactly like the deflation, then, the variance decomposes into a sum like in
eq.~\eqref{eq:variance_levels}, and also like that case, it is simple to
generalize to a hierarchical scheme by adding and subtracting larger masses.
As an identity, the extra propagators are an algorithmic crutch and need not
be related to the physical flavour content of the theory.
This achieves an additive version quite like a Hasenbusch splitting of the
fermion determinant.
It has already been noted since a long time, that the variance on the
difference is indeed suppressed~\cite{Gulpers:2014jaq}, in fact by
$a^2(m_{f}-m_{g})^2$~\cite{Giusti:2019kff}, but some additional improvements
can reduce even further the variance of both the levels.

\begin{figure}[t]
    \centering
    \includegraphics[scale=0.6]{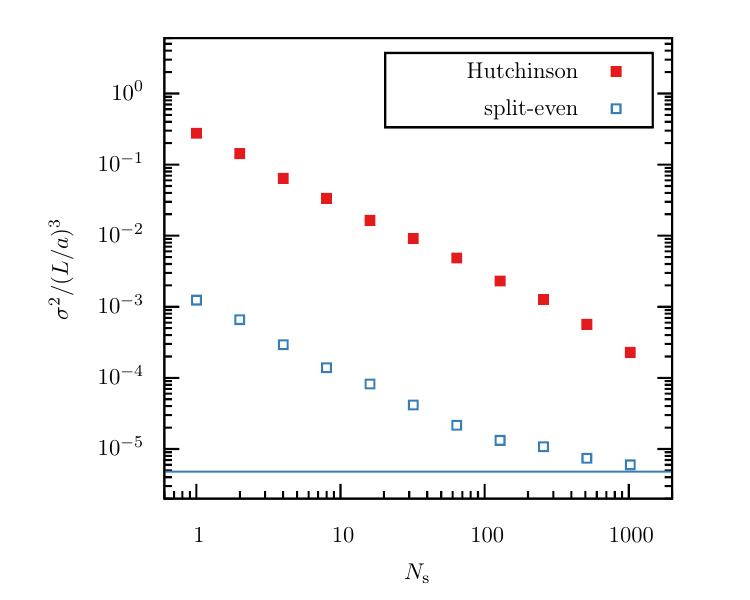}%
    \includegraphics[scale=0.6]{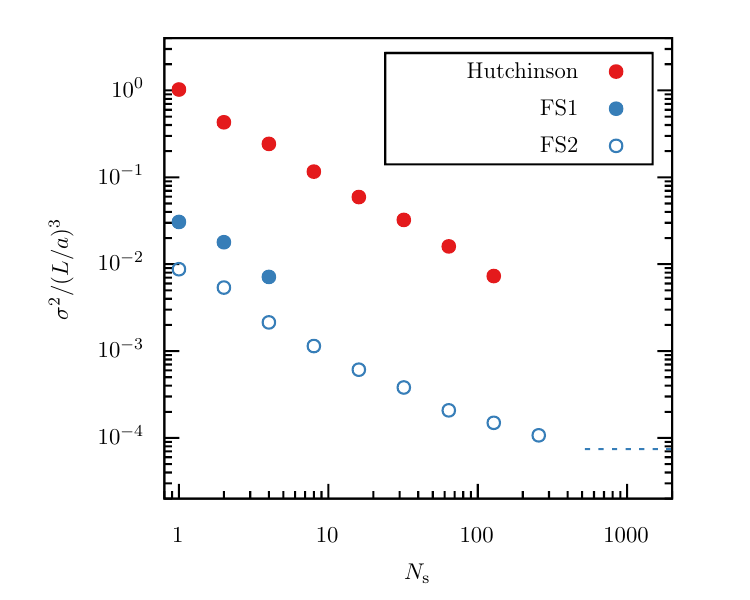} 
    \caption{The variance of the estimator for the trace of the difference of propagators
    $\mathcal S_0$ (left) or the single-propagator trace $T^{(1)}_{k,f}$
    (right). In both cases the plain Hutchinson estimator (red points) is compared with the
    improved estimators, the split-even estimator for the difference (open squares, left) and
    two variants of the frequency-splitting estimator including the hopping
    expansion (blue circles, right).}
    \label{fig:single}
\end{figure}

An improved estimator for the trace of $\mathcal S_0$ can be written with
auxiliary fields with support on the whole lattice volume, using the cyclicity
of the trace and the property $D_f-D_g=m_f-m_g$ 
\begin{align}
    \tr{\gamma_\mu\mathcal S_0(x,x)} \approx (m_{g}-m_{f})
    \frac{1}{\Ns}\sum_i (\eta_i^\dagger S_{f})(x)\gamma_\mu (S_{g}\eta_i)(x)
    \label{eq:oet}
\end{align}
which is very similar to the one-end trick used for twisted-mass
fermions~\cite{ETM:2008zte,Dinter:2012tt}, where such differences appear
naturally.
Interestingly, this so-called split-even estimator furnishes us with an
estimator for every $x_0$, which can be used to implement translation
averaging in time.
In addition, the variance is no longer independent of the gamma-matrix,
compared with the Hutchinson estimator.
We note, due to the charge factors of the electromagnetic current, in a theory
with $\Nf=2+1$ flavours, the disconnected Wick contractions result exactly in
differences of quark propagators of the light and strange quarks, and the
split-even estimator can be used by itself, and used for a multitude of
related
observables~\cite{Ce:2022eix,Boccaletti:2024guq,Djukanovic:2023beb,Djukanovic:2024cmq}.
Such a representation also works out for domain-wall
fermions~\cite{Harris:2023zsl} and has been applied to weak transition
amplitudes~\cite{Hodgson:2025iit}.

For a general flavour content, however, the single-propagator trace is
required, and an improved estimator for large quark masses can be constructed
by computing the short-distance contribution exactly via the hopping expansion
to the $k$th order~\cite{Thron:1997iy,Bali:2009hu,Gulpers:2013uca}
\begin{align}
    D^{-1}_f =  M_k + D^{-1}_f H^k
    \label{eq:hopping}
\end{align}
where $H=-(D_\mathrm{eo}D^{-1}_\mathrm{oo} + D_\mathrm{oe}D_\mathrm{ee}^{-1})$
is the hopping matrix and the sum of the first $k$ terms is given by
\begin{align}
    M_k = (D_\mathrm{ee} + D_\mathrm{oo})^{-1}\sum_{n=0}^{k-1}H^n.
\end{align}
Noting that $M_k$ is a sparse matrix, the trace of those terms can be
evaluated exactly by quadratures with so-called probing
vectors~\cite{Tang:2010}
$v^0,\ldots,v^{K-1}$ which satisfy
\begin{align}
    \sum_{n=0}^{K-1} v^n_{\alpha a}(x)v^n_{\beta b}(y) =
    \delta_{\alpha\beta}\delta_{ab}\delta(x,y) \quad\text{for all
    $\alpha,\beta,a,b,x,y$ where}\quad M_{k,a\alpha b\beta}(x,y) \neq 0.
    \label{eq:probing}
\end{align}
It turns out some fairly efficient schemes can be easily produced for small
$k$, requiring $K=24(k/2)^4$ vectors, which may be improved upon with schemes
such as those put forward in Ref.~\cite{Stathopoulos:2013aci}. 

\subsection{Numerical investigations}
\label{sec:freq-split-num}

In this section we present results for two variants of the frequency-splitting
estimator outlined in the previous subsection on the same ensemble with
$m_\pi\approx 270\,\mathrm{MeV}$ presented earlier.
{For a fair comparison, we extend the Hutchinson estimator to one which
where the sources have support on all time-slices, so that both estimators can
be used to implement translation averaging in time.}
Firstly, in Fig.~\ref{fig:single} (left) we see the variance of the first
level $\mathcal S_0$ with the second mass corresponding roughly to the
physical strange-quark mass, as a function of the number of auxiliary fields
per gauge field.
The horizontal line is an estimate of the gauge variance.
The Hutchinson estimator of the difference is shown in red solid points while
the improved split-even estimator is shown with open blue points.
The variance of the split-even estimator is $\mathrm O(100)$ suppressed with
respect to the Hutchinson estimator for the difference, which itself is only
slightly suppressed compared with the single-propagator trace shown in the
right-hand panel (solid red circles).

The right-hand panel also shows two variants of the frequency-splitting
estimator with either two (FS1) or five (FS2) levels.
The FS2 estimator uses a few levels to reach approximately a quark mass
corresponding to the physical charm-quark mass, where the hopping expansion
with $k=4$ is already efficient.
As well as the FS2 estimator having the minimum variance per cycle, we see
that it also has the minimum effective cost when looking at
Tab.~\ref{tab:cost-single}.
Optimization schemes like those of Ref.~\cite{Whyte:2022vrk}, can be useful to
further reduce the effective cost, but an order of magnitude speed-up is
fairly easy to accomplish, and appears even better at small quark
masses~\cite{Giusti:2019kff}.

\begin{table}[t]
    \centering
    \begin{tabular}{cccccccc}
        \toprule
        \multirow{2}{*}{{$m_\pi L$}} &
        \multirow{2}{*}{{estimator}} &
        \multicolumn{5}{c}{$N_\mathrm{s}$} &
        \multirow{2}{*}{$\frac{\textrm{measured cost}}{\textrm{Hutchinson $\Ns=1$}}$}
        \\
        \cmidrule(lr){3-7}
                           &            & L0  & L1 & L2 & L3 & L4 &                    \\
        \midrule
        \multirow{3}{*}{4} & Hutchinson & 100 & -  & -  & -  & -  & {100} \\
                           & FS1        & 4   & 16 & -  & -  & -  & {9.2} \\
                           & FS2        & 1   & 1  & 2  & 3  & 10 & $5.9$              \\
        \bottomrule
    \end{tabular}
    \caption{Cost to achieve a fixed variance for the single-propagator trace
        $T^{(1)}_{k,f}$ for the plain Hutchinson estimator and two variants of the
        frequency-splitting estimator.}
    \label{tab:cost-single}
\end{table}


\section{Multi-level integration}
\label{sec:multi-level}

The techniques discussed so far have concerned reducing the computational cost
of implementing translation averaging, but do not provide a solution that
deals with the vacuum contributions to the gauge variance, where they arise.
One known way to suppress those types of fluctuations is by a local update and
averaging procedure, called multi-level
integration~\cite{Luscher:2001up,Meyer:2002cd}, essentially a
generalization of the multi-hit algorithm~\cite{Parisi:1983hm}.
If both the distribution and the observable factorize, an improved estimator
can be written in which the fields in each factor in the correlation function
are averaged independently of one and another
\begin{align}
    \bar\phi^i(x) &= \frac{1}{n_1}\sum_{j=1}^{n_1} \phi^i_j(x),
    \label{eq:mlupdate}
\end{align}
where now the index $j$ labels the samples in a sub-domain keeping the
boundary field fixed, which is labelled by $i$,  harking  back to the notation of the
introduction.

\begin{figure}[t]
    \centering
    \includegraphics[scale=0.5]{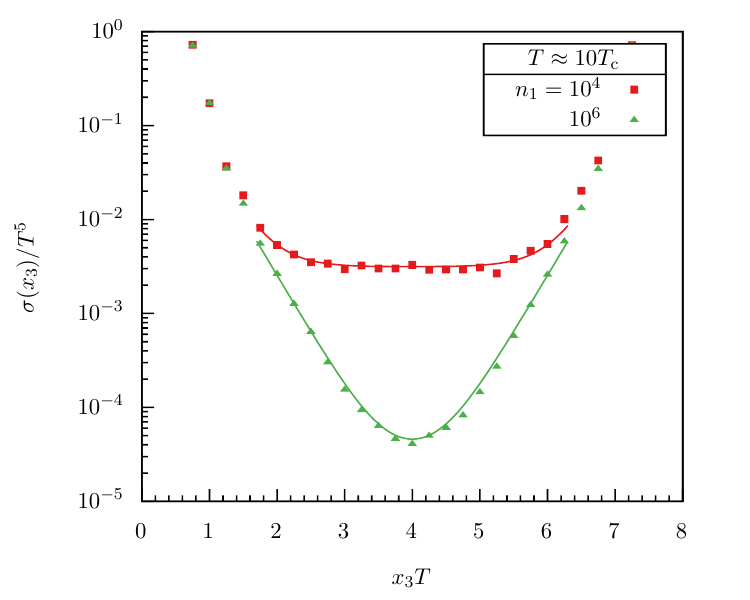}%
    \includegraphics[scale=0.5]{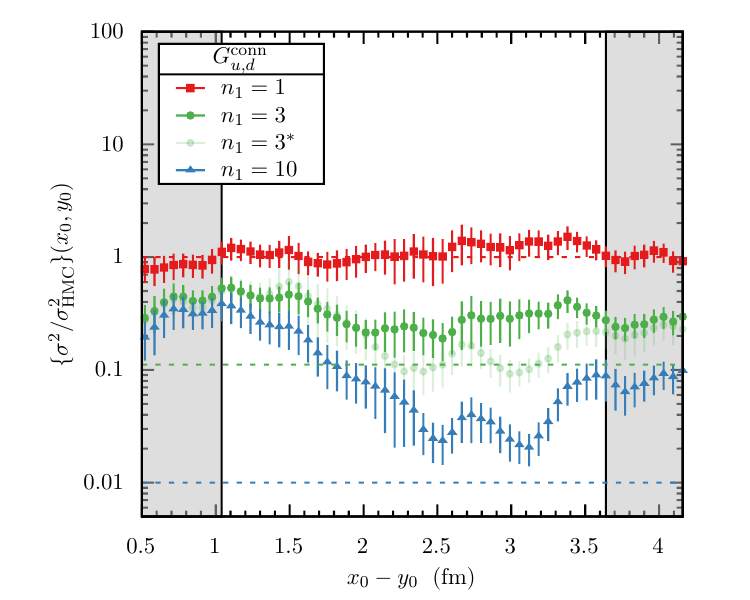}
    \caption{Variance of two-point functions with two-level integration
        schemes. The left-hand panel shows the (square root of the) variance of the spatial correlator
    of the action density for a simulation of $\mathrm{SU}(3)$ pure gauge theory
    at high temperature. The right-hand panel shows the variance of the isovector
    vector correlator using the multi-boson domain decomposed
    HMC~\cite{DallaBrida:2020cik}, compared with a
    standard HMC algorithm computed on the F7 ensemble described in the text.}
    \label{fig:ml}
\end{figure}

This results in a variance which depends on a non-trivial way on the
separation to the boundary $d$, which, if in the regime where both fields are
maximally distant to the boundary $x_0-y_0\approx 2d$, results
in~\cite{GarciaVera:2016dau}
\begin{align}
    \sigma^2 = \frac{1}{N}\Big[c_2\mathrm e^{-2m(x_0-y_0)}
        + \frac{c_1}{n_1}\mathrm e^{-m(x_0-y_0)}
        + \frac{c_0}{n_1^2} \Big],
    \label{eq:ml-variance}
\end{align}
where we note that the problematic vacuum contributions are now suppressed by
$n_1^2$, even though the cost of the simulation grows only linearly in $n_1$.
Choosing $n_1\sim\mathrm e^{m(x_0-y_0)}$ results in a constant signal-to-noise
ratio up to $x_0-y_0$.
Such an algorithm costs exponentially less than the standard update algorithm
to reach the same signal-to-noise ratio at an equivalent distance.
Fig.~\ref{fig:ml} (left) shows that the formula describes very well the
dependence of the measured variance on $x_0-y_0$ in pure gauge simulations,
where it can be applied straightforwardly.

The lack of manifest locality of the usual representation of the fermionic
path integral seems to prevent such a scheme working out in QCD.
Nevertheless, a proposal has been shown to work out in practice which
relies on the observation that an exact factorization is not required to be
profitable~\cite{Ce:2016idq,Ce:2016ajy}.
If a decomposition exists such that one term (the approximant) is
factorizable, while the remainder has a small variance (similar to the
decompositions presented for the quark propagator in the previous sections)
then a large acceleration can be achieved.
The scheme put forward by C\`e, Giusti and Schaefer is based on an overlapping
parallel Schwarz method, related to the method for solving classical partial
differential equations, and the multi-boson
representation~\cite{Luscher:1993xx}.
A complete exposition of the factorization scheme is beyond the scope of these
proceedings, but recent results for such update schemes for the fermion
determinant show that indeed the fluctuations in the non-factorized remainder
are under control, and that the variance has the expected
suppression~\cite{DallaBrida:2020cik}, see Fig.~\ref{fig:ml} (right).
Recent progress was also shown at this conference by Barca et
al.~\cite{Barca:2025dca}.
These methods are a promising way to ameliorate the signal-to-noise ratio
problems discussed in the opening section and in particular may unlock
ground-state matrix elements needed for precision studies of the Standard
Model.


\section{Conclusions}
\label{sec:concs}

In these proceedings, I have reviewed the origins of signal-to-noise ratio
problems in lattice QCD, attempting to paint a picture of some obstacles to
precision computations of observables.
Translation averaging is one of the simplest ways to reduce the variance but
it does not come with an acceptable cost for quark propagators if implemented
in a naive fashion.
I have presented a selection of methods to accelerate the computation of quark
propagators for large and small separations, based on decompositions of the
propagator and the observable of interest.

For large separations, the quark propagator is well represented in a subspace
created from blocked low modes of the Dirac operator, leading to a multigrid
low-mode averaging scheme.
Furthermore, the little operator is well conditioned if the subspace spans
both chiralities of the fields, leading to an efficient deflation scheme
which has a favourable scaling with the physical volume.
For the quark propagator at short distances, instead, a good approximation is
constructed by shifting the quark mass and using the hopping representation
for the largest masses.
Noting that the sum of the first few hopping terms is a true sparse matrix, at
least in the Wilson formulation, enables its trace to be computed exactly with
a few probing vectors.

Nevertheless, the vacuum contributions to the variance cause severe
signal-to-noise ratio problems that are not solved by translation averaging.
A local update procedure such as multi-hit or multi-level integration
provides a clear route to improve the computation of $n$-point
functions, and recent innovations have shown that an exact factorization is
not necessary for this idea to be useful.
Empircal results have shown that the multi-boson
domain-decomposed HMC allows one to compute a local average for the
contribution which suffers most of the fluctuations, and therefore can speed
up the computation exponentially for large separations.
To attain the next level of precision in hadronic matrix elements, some
combination of the methods presented here should be beneficial, and progress
on multiple fronts suggests that significant improvements may be within reach.


\section*{Acknowledgments}

I extend my sincere thanks to the organizers of Lattice 2025 for a 
scientifically illuminating program and a wonderful conference.
I appreciate all of the valuable discussions with members of the lattice
community I have had on topics related to these proceedings, and am especially
grateful to Mattia Dalla Brida, Leonardo Giusti, Maxwell T.~Hansen, Harvey
B.~Meyer, Agostino Patella, Mike J.~Peardon, my colleagues in the RC$*$
Collaboration and Marina K.~Marinkovi\'c for encouraging work in this
direction.
I am indebted to Roman Gruber for his persistence in the investigations of
multigrid low-mode averaging schemes.

\printbibliography

\end{document}